# Emerging Technologies for Cancer Research: towards Personalized Medicine with Microfluidic Platforms and 3D Tumor Models

Matteo Turetta[a,†], Fabio Del Ben[b,†], Giulia Brisotto[b,c], Eva Biscontin[b], Michela Bulfoni[a], Daniela Cesselli[a], Alfonso Colombatti[d], Giacinto Scoles[e], Giuseppe Gigli[e,f], Loretta L. del Mercato[e,*]

[a]*Istituto di Anatomia Patologica, Dipartimento di Scienze Mediche e Biologiche, Azienda Sanitaria Universitaria Integrata di Udine, 33100 Udine, Italia;* [b]*Patologia Clinica Oncologica, Oncologia Sperimentale 2, Dipartimento di Ricerca Traslazionale, CRO Aviano I.R.C.C.S., 33081 Aviano (PN), Italia;* [c]*Istituto Oncologico Veneto IOV, I.R.C.C.S., 35128 Padova, Italia;* [d]*Oncologia Sperimentale 2, CRO Aviano I.R.C.C.S., via F. Gallini 2, 33081 Aviano (PN), Italia;* [e]*Istituto di Nanotecnologia, CNR-Nanotec, Polo di Nanotecnologia c, 73100, Lecce, Italia;* [f]*Dipartimento di Matematica e Fisica "E. De Giorgi", Università del Salento, 73100 Lecce, Italia*

**Abstract:** In the present review, we describe three hot topics in cancer research such as circulating tumor cells, exosomes, and 3D environment models. The first section is dedicated to microfluidic platforms for detecting circulating tumor cells, including both affinity-based methods that take advantage of antibodies and aptamers, and "label-free" approaches, exploiting cancer cells physical features and, more recently, abnormal cancer metabolism. In the second section, we briefly describe biology of exosomes and their role in cancer, as well as conventional techniques for their isolation and innovative microfluidic platforms. In the third section, the importance of tumor microenvironment is highlighted, along with techniques for modeling it *in vitro*. Finally, we discuss limitations of two-dimensional monolayer methods and describe advantages and disadvantages of different three-dimensional tumor systems for cell-cell interaction analysis and their potential applications in cancer management.

**Keywords:** Cancer, circulating tumor cells, exosomes, tumor microenvironment, 3D cell cultures.

## 1. INTRODUCTION

Despite huge investments for decades in oncology research, cancer is still among the leading causes of death worldwide and its death toll is expected to rise by about 70% over the next two decades [1]. There are many aspects of cancer, including its heterogeneity, complexity, and dynamic nature, which require a radical change in the way we approach both cancer study and its management. For instance, the tumor genetic profile and its microenvironment play a critical role in cancer development and progression and could affect patient treatment response [2,3]. Recent evidence shows that matching the patient tumor profile to treatment protocols yield better results in terms of outcome. This approach has been called personalized medicine [4–7], however, in order to pursue this model of medicine, novel techniques capable of earlier cancer detection, intensive and minimally invasive therapy monitoring and drug development are needed.

Core needle biopsy is a standard procedure in oncology for cancer diagnosis and treatment planning, however it suffers from several limitations.[8,9] Tumor heterogeneity is a key feature of cancer and occurs both between cancer cells of a single tumor (intra-tumor heterogeneity) and between tumors of the same type across different patients (inter-tumor heterogeneity). The small amount of biological material obtained from biopsy might not be sufficient to reflect the genotypical and phenotypical heterogeneity of the disease [10], moreover, core needle biopsies of masses located in delicate or hard-to-reach organs, such as lung, kidney, and brain, are risky and rarely repeatable [11]. A potential innovation in this field is the so-called *liquid biopsy,* that is the analysis of cancer biological material, such as circulating tumor cells (CTCs), cell-free circulating tumor DNA (ctDNA) and circulating-tumor derived exosomes, released into the peripheral blood from the primary tumor and metastasis. This approach has great potential to revolutionize the current clinical practice, by providing easy and repeatable access to the heterogeneous tumor biological material, and consequently to the information about disease state, prognosis and chemo-sensitivity.

On another side, the implementation of suitable *in vitro* tumor models for cancer and microenvironment studies, capable for example of predicting, for each patient, the response to specific chemotherapeutic agents represents another challenge of personalized medicine. Indeed, traditional two-dimensional *in vitro* models frequently fail in predicting the *in vivo* efficacy of anticancer therapies and are being replaced by three-dimensional (3D) systems that better mimic the *in vivo* behavior of cells in tumors.

In the present review, we describe three hot topics in cancer research. The first section is dedicated to new microfluidic techniques to be implemented in '*liquid biopsy*' for CTCs detection. The second section is focused on describing the available devices for exosome investigations and summarize the clinical evidences that support their potential clinical role. In the last section, we will highlight some of the most popular examples of 3D cell cultures currently used, including spheroids, hydrogels, and fibrous-material-based scaffolds.

## 2. MICROFLUIDIC PLATFORMS FOR SINGLE CELL-ANALYSES

Current standard of care for cancer diagnosis relies on "solid" biopsy, which, however, presents several limitations, as it requires invasive, risky and painful procedures, rarely repeatable and in some cases challenging due to the remote tumor localization. Moreover, the small amount of tissue obtained through a single biopsy could be insufficient to characterize intratumor heterogeneity, providing misleading or insufficient information for treatment decisions. Biopsy at metastatic sites is extremely uncommon [12], thus failing to represent biological evolution of the disease and the subclones driving metastasis. These limitations have led to an intensive effort on the quest for new cancer biomarkers in body fluids (including peripheral blood, urine, stools, cerebrospinal fluid, tears and saliva) leading to the recent diagnostic concept of '*liquid biopsy*', which consists of the analysis of tumor biological material, such as CTCs, ctDNA, and exosomes.

### 2.1. Overview and role of CTCs in cancer management

CTCs refer to the cells shed from either the primary tumor or metastatic site into the bloodstream. These cells hold relevant information about cancer progression and metastasis, both in their number and trend over time [13], and in their biological features [14]. Therefore, capture of CTCs from the bloodstream and their characterization hold great promise to improve cancer diagnosis and treatment [15].

Currently, the presence of CTCs in the peripheral blood above a set threshold, as detected by the CELLSEARCH® CTC Test, is associated with decreased progression-free survival and decreased overall survival in patients treated for metastatic breast, colorectal, or prostate cancer. The test is FDA-cleared as an aid in the monitoring of patients of said cancer types, at any time during the course of disease, allowing assessment of patient prognosis and prediction of progression-free survival and overall survival.[16–18]

Nevertheless, testing for CTCs is currently not widely adopted in the clinical routine. Among drawbacks of CELLSEARCH® are the relatively high cost per analysis and low sensitivity (e.g.: 61% in metastatic breast cancer [13]), but the strongest reason preventing its widespread adoption in the clinic, in our opinion, is that the clinical utility of testing for CTCs is still controversial. A big role in the debate is played by the failure of the Phase III SWOG S0500 trial, which had the objective of investigating the benefit of an early switch of chemotherapy guided by the number of CTCs. The design of the study, as commented by Raimondi and colleagues [19], had some pitfalls (part of therapies left unchanged between arms, lack of classification according to molecular subgroups), which could have messed up a clean benefit demonstration. In opposition to the S0500, the only trial showing the ability of CTCs to drive a therapy has been published by Scher et al.[20] This trial successfully demonstrated that, in patients with metastatic castration-resistant prostate cancer, a therapy choice based on CTCs can improve the survival of patients. Specifically, the characterization of a protein expression (AR-V7) in the nucleus of CTCs is a treatment-specific biomarker that is associated with superior survival on taxane therapy over Androgen receptor signaling-directed therapy.

Thus, both the number and the characterization of CTCs showed to carry significant information potentially impacting the metastatic cancer patient management. While clinical validation of CTCs as prognostic and predictive biomarkers is out-of-discussion, their clinical utility is only emerging and needs stronger evidence to be fully supported by the clinical community.

### 2.2. Microfluidic techniques for CTCs detection and isolation

Detecting CTCs is challenging, because they occur at very low concentrations, around a single tumor cell in a background of a billion blood cells [21]. Therefore, their identification and characterization requires methods of extremely high analytical sensitivity and specificity, which usually consist of a combination of enrichment and detection procedures [11].

During the past decade, microfluidics devices have emerged as powerful tools for both basic and applied research on cancer. This technology offers the possibility to precisely control small volumes of fluids (down to a pico-liter), by using channels with dimensions of ten to hundreds micrometers, and to simultaneously handle multiple samples in multiple bioreactors [22]. Among the several possible approaches for fabricating microfluidic devices, soft-lithography and poly-dymethylsiloxane (PDMS) have become the most widely represented in academia for biological applications [23]. This is due to several properties of PDMS such as flexibility, allowing relatively easy and rapid fabrication of devices with various types of channel geometry [22]; transparency, providing excellent live cell imaging conditions; gas permeability, essential for cell survival. Therefore, the field of microfluidics offers several essential advantages including reduced sample volume and reagent consumption, fast processing speed, low cost, high sensitivity and enhanced spatial and temporal control, highlighting its clear potential to advance cancer research in a new and unconventional way. In this context, microfluidics is a suitable tool for analyzing complex fluids *in vitro,* and several emerging microfluidic approaches can isolate CTCs, exploiting their biological or physical properties (Fig. **1**), thus potentially impacting in cancer diagnosis and management [24]. For the purpose of the present review, we will focus on microfluidic techniques for CTCs detection and isolation.

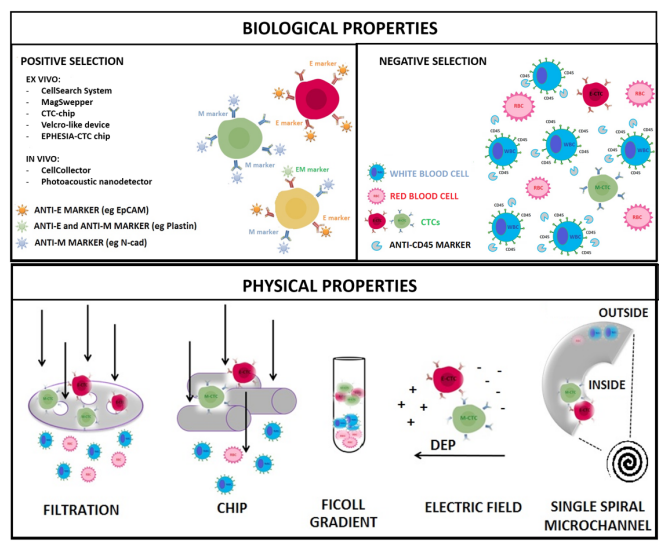

**Fig. (1).** Schematic view of technologies for CTCs isolation. Methods based on biological properties, described in §2.2.1 and on physical properties, described in §2.2.2.

*2.2.1 Affinity-based methods*

One of the most widely used methods for isolating CTCs is based on the affinity of a specific antigen expressed on the cell surface to its corresponding antibody, typically bound to either a device surface or a magnetic particle. In this way, CTCs are trapped on the device surface while most of the undesired blood cells are flowed away. This principle has been widely employed for the detection of CTCs, as it offers high specificity of the recovered CTCs and improves the isolation purity [18].

CTCs have been extensively detected using the epithelial adhesion molecule (EpCAM), a tumor specific marker commonly expressed by cancer cells of epithelial origin. An example is the CTC-chip, which consists of EpCAM-functionalized microposts for capturing CTCs from whole blood with high sensitivity and purity [25]. Following this, a number of technologies with geometrically enhanced microstructures were developed, aiming at increase the degree of sensitivity and purity of CTCs isolation. In this context, the Herringbone chip or "HB-chip" was developed by using microchannel fabricated into a herringbone shape to disrupt streamline flows, which may reduce the chance of interaction between target cells and the antibody coated surface [26]. Another example is the GEDI chip, which increased the capture efficiency by optimizing the displacement, size and shape of microposts. Moreover, capturing of CTCs was achieved by using another antibody than EpCAM, namely a prostate specific membrane antigen (PSMA). A further CTC-chip based system is the commercially available CEE$^{TM}$ microfluidic chip, characterized by randomly positioned and streptavidin-functionalized microposts, which allow capturing CTCs tagged with biotinylated antibodies [27]. The NanoVelcro CTC chip is based on silicon nanowire substrates (SiNWS) coated with EpCam for CTCs capture [28,29].

The major downside of positive selection is the *a priori* selection of a specific subpopulation of CTCs, thus lowering the sensitivity of the test and missing potentially important subpopulations. EpCAM, for example, includes only cells of epithelial phenotype, while mesenchymal cells are excluded from the detection. Mesenchymal and stem-cell like CTCs have proven to be strong indicators of disease progression, and attention should be paid not to overlook them [30,31].

An alternative approach to positive selection, in which tumor-specific markers are used to target CTCs, negative selection methods have also been explored based on the removal of cells of hematopoietic origin by targeting antigens mostly not expressed by CTCs, such as CD45 [21]. Recently, Ozkumur *et al.* developed the CTC-iChip aimed at isolating CTCs by removing blood cells on the base of physical properties and CD45 and/or CD15 expression [32]. Another example is the geometrically activated surface interaction (GASI) chip, which was similar to the HB-chip but with enhanced microvortexing properties aimed at increasing the number of captured leukocytes [32]. Overall, negative selection allows the capture of CTCs with less or no expression of EpCam and the eluted CTCs can be collected intact and viable [33].

Negative selection might, as a downside, exclude circulating tumor cells, as suggested by the findings of double positive CK+/CD45+ [34]. Furthermore, one of the first studies on CTCs detection chose to exclude CD45 from the discriminating criteria to detect CTCs [35].

Thus, caution should be used in considering all CD45+ events as white blood cells, since CTCs might show expression of CD45 comparable to some leukocytes populations, especially non-lymphocytes populations (low expression of CD45).

As an alternative to the traditional antibodies, recently, aptamers have gained great attention, as they have similar functions but more advantages for biological applications. Aptamers are defined as small oligonucleotides, such as RNA or DNA, or peptides, which can bind with high specificity and sensitivity to a variety of molecular or cellular targets. They can be easily synthetized and selected by an *in vitro* process known as Systematic Evolution of Ligands by EXponential enrichment (SELEX) (Fig. **2**).

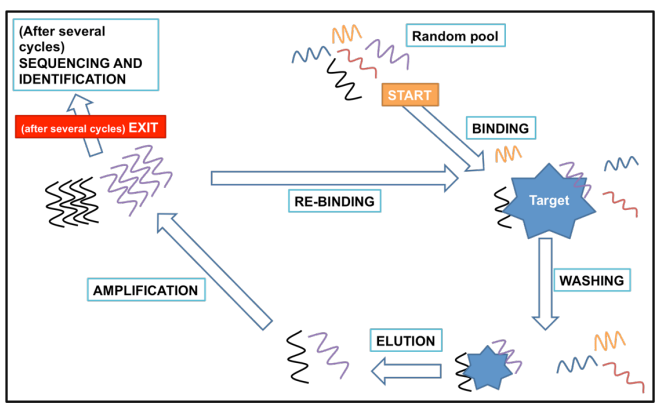

**Fig. (2).** Schematic view of the SELEX technology. Starting point is a random pool of synthetic DNA oligonucleotides. The SELEX procedure consists of several cycles of the following steps. At first, the random pool and the target molecules are incubated for binding. Unbound oligonucleotides are then removed by washing. The target-bound oligonucleotides are eluted and amplified by PCR or RT-PCR. A new pool of oligonucleotides is thus generated, and is then used for the next selection round. In general, 5 to 20 complete cycles are needed for the selection of specific aptamers. The process can be interrupted and selected oligonucleotides can be characterized by sequencing.

The key benefit of applying aptamers for the isolation of CTCs is that they can be prepared in panels targeting multiple proteins expressed on available cancer cells, without necessarily knowing the precise targets. The potential of using aptamers for the enrichment of CTCs has been shown with artificial samples prepared by spiking several cancer cell lines (e.g. leukemia, colon cancer and glioblastoma) in whole blood samples [36–38]. Fang's group has generated a panel of aptamers through *in vitro* cell-SELEX process against a non-small cell lung cancer (NSCL) cell line [39] and patterned the NanoVelcro Chip with two of these aptamers, which allowed not only to capture but also to recover the NSCL cells by using a nuclease solution [40]. Recently, they demonstrated a rational design for an aptamers cocktail grafted onto the NanoVelcro Chip, highlighting the synergistic effect of aptamers in the recognition of cancer cells with respect to the single aptamer based method. This enhanced and differential capture performance was demonstrated both in experiments with cell lines and in patient samples [41].

### 2.2.2 "Label-free" methods

Label-free methods allow isolation of cells untouched by antibodies, which can affect the viability of the cells or their gene expression patterns [42]. Moreover, monoclonal antibodies are quite expensive and require a priori knowledge of the target protein, while cancer is by definition a transforming, dynamic disease and cells are frequently in transition between epithelial, mesenchymal and stem cell state, with frequent protein expression rearrangement [43]. Thus, targeting a particular protein(s)expression is truly not the most comprehensive way to dissect the heterogeneity of CTCs. Intact cell isolation is crucial to perform further purification or analysis of their morphological or biological properties.

Biophysical separation approaches are part of the so-called label-free methods, since they do not require the use of antibodies directed versus specific antigens [21]. Instead, they rely on physical characteristics such as density, size, and electric charge to discriminate between CTCs and other cells (e.g., leukocytes). This conceptually overcomes misdetection of CTCs lacking a specific protein.

*Density-based gradient or isopycnic density gradient centrifugation* is a robust technique used as a first CTCs enrichment step. This strategy grounds on the differential migration of cells according to differences in buoyant density. The centrifugation of blood in the presence of Ficoll or Oncoquick, two commercially available solutions, allows to separate CTCs from erythrocytes and granulocyte [44]. Density-based approaches result in an insufficient purity for most downstream analyses and typically require further enrichment steps. Their main limitation is a non-specific loss of target cells, due to the presence of CTCs with density comparable to white blood cells [44].

*Dielectrophoresis* (DEP) is an innovative approach to cell separation that exploits the distinct electrical fingerprints of different cells and is based on the movement of neutral polarizable particles induced by electric field gradients [21]. ApoStream, for example, is a commercial system for the enrichment of CTCs, adopting this strategy to effectively isolate CTCs from clinical samples [45,46]. While this strategy recognizes cells based on their electrical signature, DEP can also be applied as a technique to finely manipulate single-cells, recognized by other means. In this context DEP is not a label-free approach, rather a mere technique to move cells, as in DEPArray system (Silicon Biosystems), which traps single cells in DEP cages generated via an array of individually controllable electrodes, and is able to gently manipulate and sort single, viable cells by a multiparametric fluorescence-based selection process. This system is designed for single-cell recovery suitable for downstream analyses. In fact, multiple clinical studies have used DEPArray to recover single CTCs for subsequent genetic analyses following enrichment using centrifugation or positive or negative selection (i.e., CellSearch or CD45 immunomagnetic depletion) [43,47–51]. Several clinical settings have been explored, such as primary and metastatic breast cancer, metastatic lung and colon cancer. However, technologies based on DEP have limited selectivity and throughput [52].

*Membrane-based filtration* is one of the first technologies used to isolate CTCs based on cell size and deformability. Several methods were developed, like the commercially available technology ISET (Isolation by Size of Epithelial Tumor cells). In principle, the 8 µm-pores of the polycarbonate membrane-filter should retain larger CTCs while smaller leukocytes should pass through [53]. Microfiltration allows rapid processing of blood for the enrichment of CTCs, however, these systems are prone to clogging, which decreases enriched sample purity and standardization of the process. Some setups require parallel processing with multiple filters for large volumes [51,54]. Size-based separation is limited by large variations in the size of CTCs, depending on the tumor type and within the same patients, and typical capture purities are less than 10% [21]. The major limit of label-free technologies based on size is that some CTCs and leukocytes sizes are significantly overlapping. For example, in metastatic prostate cancer, very small CTCs have been detected [30,55], and CTCs in aggregates (or clusters) have a smaller morphology than individual, non-aggregated CTCs [56]. This variability on size can cause cell loss and low purity [57].

Methods preserving the integrity of clusters of CTCs are interesting because these clusters are believed to have more metastasizing capability than single cells, because they represent a heterogeneous mix of cancer cells.

Hydrodynamic-based methods have the highest throughput capability. In microchannels, fluid shear generates lateral forces, which cause focusing and inertial migration of particles, that can achieve a high throughput separation based on size. These techniques exploit three important forces that move the flowing particle to equilibrium position according to its density (shear-induced, stresslet-velocity-field-induced and wall lift). In curved channels, vortexes or secondary flows are superimposed to primary ones, and defined by a non-dimensional number called Dean number [58]. By tuning the dynamic equilibrium between these forces, CTCs can be isolated from other cells. Regardless of the type of technology, the goal is to obtain different flow velocities based on cell size differences to separate CTCs with high efficiency. Several architecture and different methods were proposed, for example, p-MOFF (parallel-Multiorifice Flow Fractionation) used a series of contraction and expansion structures in parallel to trap CTCs from 24 breast cancer patients [59]. The ultrahigh-throughput spiral exploit the inertial focusing inside this multiplex spiral microfluidic device to rapidly separate rare cells from a large volume of blood samples (7.5mL in 10 minutes) in label-free manner. This device allows downstream analysis such as fluorescence *in situ* hybridization, FISH [60]. The Vortex device, together with an in-flow automated processing system, can release viable CTCs in a small volume (300 μL). which can undergo standard assays downstream, such as cytology and cytogenetics (Papanicolaou staining, FISH) and can be useful for genomic profiling, transcriptomic and drug assays. The enumeration is label-free: cell can be counted in-flow, without any cell-specific label and remain unaltered, allowing collection off-chip for further assays [61].

Recently, our group has exploited metabolism to detect CTCs [62]. Metabolic reprogramming is considered one of the hallmarks of cancer [63], and is the working principle of current gold-standard in cancer imaging, Positron Emission Tomography (PET). In fact, PET-scan relies on the increased uptake of glucose by cancer cells with respect to normal cells. As an added abnormal metabolic feature, cancer cells are known to generate acidity in the surrounding microenvironment, with evidence pointing at two main causes: abnormal glucose metabolism (Warburg effect) and ion-channels abnormal expression (NHE Sodium-proton antiporter) [64–66]. The Warburg effect consists of increased glucose consumption and lactate production even in the presence of oxygen, with lactate extruded from the cells together with protons, thus causing a decrease in the pH of tumor extracellular environment. NHE overexpression leads to increased efflux of protons and consequent pH drop outside the cells [64]. The extreme rarity of CTCs, though, brings the alteration of pH in a whole blood sample to undetectable levels. Our group addressed this problem using microfluidics to emulsify a blood sample in millions of monodisperse pL droplets containing single cells, together with a fluorescent pH-probe. Once the emulsion is produced, droplets can be reinjected to detect and quantify in real-time their pH value by laser-induced fluorescence (Fig. 3) [62].

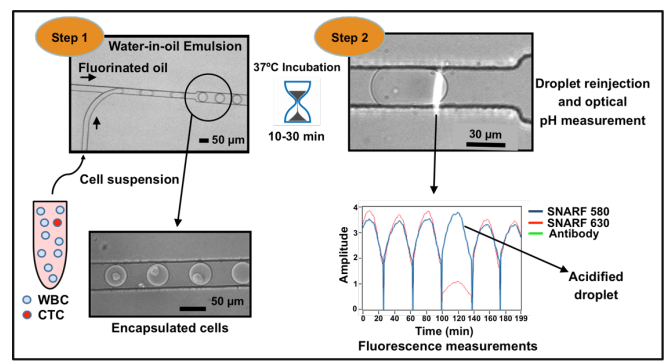

**Fig. (3)**. Schematic representation of metabolism-based method steps. **Step 1:** starting from a cell suspension in which both white blood cells (WBCs) and cancer cells (CTCs) are present, the sample, containing a fluorescent pH probe (e.g., SNARF), is emulsified in millions of picoliter droplets at a microfluidic flow-focusing junction using fluorinated oil and surfactant to avoid droplet fusion. Encapsulation of single cells can be seen at bottom-left corner. Then, the emulsion is incubated for a variable amount of time at 37°C to activate metabolic processes. **Step 2:** Droplets are reinjected and fluorescence is screened with a laser slit to measure pH. In the bottom-right corner a sample track is displayed, showing the altered emission ratio between two fluorescent channels (SNARF 580 and SNARF 630) indicating an alteration of physiological pH [62].

CTCs are thus detected by specific changes in pH or lactate concentration without the need for surface-antigen labeling [62]. With this method tumor cells from several cell lines could be detected in spiked samples and putative CTCs could be enumerated in exploratory clinical trials. Notably, besides single cells, cell clusters in the blood of some breast cancer and lung cancer patients could be observed. Further work is on going to clarify the clinical validity of the method, by evaluating the performance of the number of metabolically active CTCs as a surrogate of survival, prognostic and predictive indicator. The integration of multiple approaches might enhance the capture efficacy of CTCs, purity, viability and yield with sufficient throughput [67].

It has to be stressed that some of the above-described technologies have been tested only with cancer cell lines, but not with patient specimen. The use of cancer cell lines as a model of CTCs could be a convenient tool for proof of concept and analytical validation, but it fails to represent the complexity of ex vivo samples, thus being unfit for clinical validation of the technology. With respect to size, for example, some CTCs are smaller than cancer cell lines and more deformable; with respect to protein markers, cell lines are rather homogeneous, in contrast to heterogeneity of "real" CTCs [43].

## 3. MICROFLUIDIC PLATFORMS FOR EXOSOME ANALYSES

### 3.1. Exosomes and their role as a cargo carriers

Exosomes are heterogeneous phospholipid cell-derived nanovesicles (30–100 nm in diameter) released by a variety of cells. They are generated inside multivesicular endosomes or multivesicular bodies (MVBs) and, upon fusion with the plasma membrane, they are released into the extracellular environment [68]. Analysis of their protein composition confirmed that they are actively secreted by living cells and commonly express tetraspanins, a class of membrane proteins including CD9, CD63 and CD81 [69]. Other frequent exosomal proteins, reported in the ExoCarta database, are GTPases, cytoskeletal proteins, Caveolin-1, Ras proteins, annexines (I, II, V and VI), ANXA proteins, Albumin, Alpha-enolase, Glyceraldehyde 3-phosphate dehydrogenase, CD146, Milk Fat globule-EGF-factor VIII, Tyrosine 3-monooxygenase/tryptophan 5-monooxygenase activation protein zeta polypeptide, Pyruvate kinase, CD18, CD11a, CD11b, CD11c, CD166, CD58 and the Heat Shock Proteins (Hsp70 and Hsp90), which act as facilitators to load peptides for the major histocompatibility complex I [70]. Exosomes are able to protect and deliver to target cells a variety of macromolecules including proteins, mRNA, miRNA and DNA (Fig. 4) [71]. Proteome and transcriptome expressed in such vesicles can often be considerably different from what expressed from the cell of origin, since exosomes are produced by an active sorting mechanism. Moreover, specific miRNA can be encapsulated and then transferred to receiving cells, altering their gene expression and thus demonstrating functional effects [72].

Several studies on exosomal miRNAs have given proof of their role in stem cell differentiation, organogenesis (miR-1) and hematopoiesis (miR-18), as well as in tumorigenesis (miR-17, miR-18, miR-19a, miR-20, miR-19b-1, miR-93-1) and finally in the process of metastasis [73–78].

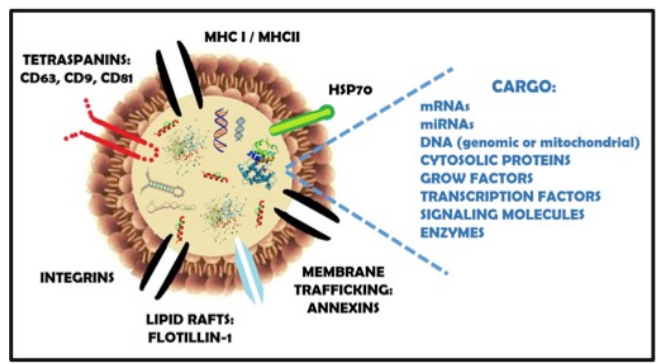

**Fig. (4).** Exosome structure and content. Exosomes are small membrane bound vesicles characterized by a phospholipid bilayer and by the presence of proteins involved in membrane trafficking (e.g. annexins and flotillin), cell targeting (e.g. tetraspanins and integrins) as well as other proteins involved in exosomal biogenesis (Alix and TSG101, not shown). Additionally, depending on the cell of origin, exsosomes contain a molecular cargo constituted by cytosolic proteins, growth factors, cytokines, DNA, RNA and miRNA, which can be delivered to target cells. Abbreviations: HSP, heat shock protein; MHC, major histocompatibility complex.

### 3.2. Role of exosomes in cancer

Recently, exosomes have emerged as a new promising class of circulating biomarkers, especially in oncology, because of their presence, at very high concentrations, in most biological fluids such as urine, amniotic fluid, malignant ascites, bronchoalveolar lavage fluid, synovial fluid, breast milk, saliva, cephalorachidian fluid and blood [79]. They play various roles in influencing the tumor microenvironment, modulating the immune response, regulating the intercellular communication and in the process of tumor resistance by mediating drug efflux [80–82].

It has been demonstrated that the increase of specific exosomal markers, such as the tetraspanins CD9, CD63 and CD81, can be used in the diagnostic procedures of several tumors and infectious diseases. For example, it has been reported that CD63 increases in the plasma of patients affected by melanoma [83] and other tumors [84], while CD81 is significantly higher in the serum of chronic hepatitis C patients [85]. Similarly, increased concentration of activating transcription marker 3 and Fetuin-A has been identified in urinary exosomes of patients suffering from renal diseases, [86]. Other exosomal molecules present in urine can be used as either bladder or prostate tumor markers. In this regard, PCA-3 and TMPRSS2 have been suggested for the diagnosis and monitoring of prostate cancer [87,88].

Similarly, TGFβ has been detected in the exosomes isolated from the serum of patients affected by brain tumors, while exosomes positive for the Epithelial Growth Factor Receptor vIII (EGFRvIII) have been proposed as a specific marker of glioblastoma bearing the EGFRvIII mutation [89]. Glioblastoma microvesicles transport RNA and proteins that promote tumor growth and provide diagnostic biomarkers [90].

Glypican-1 (GPC1)-positive exosomes have been detected both in breast and pancreatic cancer patients suggesting that GPC1 can be a pan-specific biomarker for cancer exosomes [91]. Importantly, in pancreatic patients GPC1 exosomes are prognostic [91].

As said before, miRNAs encapsulated within exosomes can be transferred to recipient cells altering their gene expression and thus mediating functional effects. For this reason, their analysis may be included in the diagnostic strategies of several tumors. In fact, exosomal miRNA levels were reported to be increased in the serum of patients affected by lung cancer, suggesting their potential as a useful tool for the diagnosis of lung adenocarcinoma [92,93]. Additionally, elevated serum levels of exosomal miR-141 and miR-375 were found to correlate with disease progression in prostate cancer patients, while miR-21 and miR-1246 were associated, in esophageal squamous cancer, with drug resistance and poor prognosis, respectively [94,95]. Regarding the ovarian cancer, it has been reported that 8 circulating exosomal miRNAs are relevant for distinguishing benign from malignant diseases [96].

In conclusion, exosomes and their associated molecular cargo may be used as new potential diagnostic and prognostic biomarkers. However, the major barrier in supporting their potential clinical use is the optimization of standardized methods to isolate, quantify and molecularly characterize these nanosized extracellular vesicles.

### 3.3. Methodological principles for the isolation of exosomes.

Several methods for exosomes isolation have been developed in the research field during the past three decades, including ultracentrifugation, chromatography, filtration, polymer-based precipitation and immunological separation (Fig. **5**).

However, at present, none of them is suitable for clinical applications. In fact, due to the biological complexity and variability among body fluids, the isolation of exosomes appears extremely challenging. The most critical factor is the contamination from non-exosomal material recovered during the procedures for the isolation of these tiny nanovesicles: all developed procedures are variably able to enrich the exosomal fraction, but fail to produce pure exosome preparations. Exosomes' enrichment may also be influenced by a pre-processing delay, by the chosen separation method and, finally, by the presence of contaminants, such as fibrinogen and albumin, in the case of blood samples.

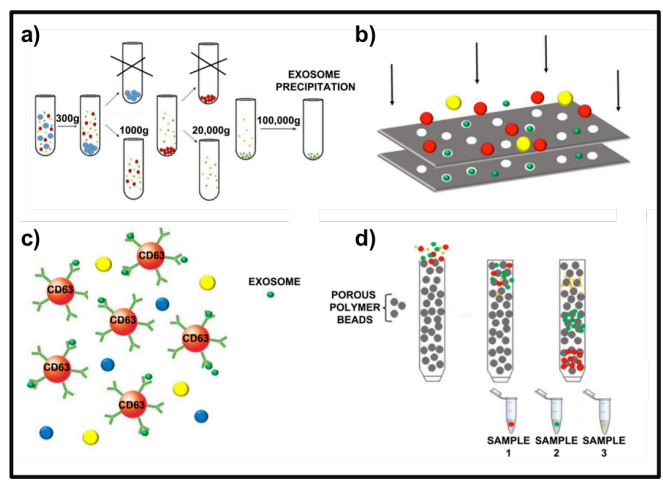

**Fig. (5).** Methods for exosome enrichment from biological fluids. Representative pictures of **(a)** differential ultracentrifugation protocol; **(b)** size-based separation chromatography; **(c)** immunoaffinity approach for antibody-specific capture; **(d)** polymer-based precipitation technique.

Another major issue, connected to their small size, is that exosomes are below the field of detection for most analysis methods presently available. As a consequence, recovery and contamination cannot be reliably quantified and standardized isolation protocols have not been developed yet. An important largely accepted statement is that neither the size nor the morphology nor the biological composition are valid criteria for distinguishing and characterize these nanovesicles [97].

Differential ultracentrifugation is actually considered the "gold standard" isolation method, where the centrifugal force is used for nanoparticle precipitation [98]. The major issue, encountered if high centrifugal acceleration is applied, consists in the fusion of vesicles and contamination of pellets with proteins or free miRNAs as well as a possible change in their functional properties [99]. To reduce the co-precipitation of protein aggregates and other particles in the exosome fraction, most recent studies suggest a combined approach of ultracentrifugation and sucrose density gradient [100].

A commercial kit, available for more than 10 years, ExoQuick™ (System Biosciences, Mountain View, CA, USA), represents an easy and fast polymer-based precipitation technique for exosomes isolation (Fig. **6**). It works just adding the solution to the biological fluid. Its disadvantages are the contamination with non-exosomal materials and the interference of the polymer during the downstream analyses. With respect to ultracentrifugation, polymer precipitation yields an increased concentration of exosomes [97].

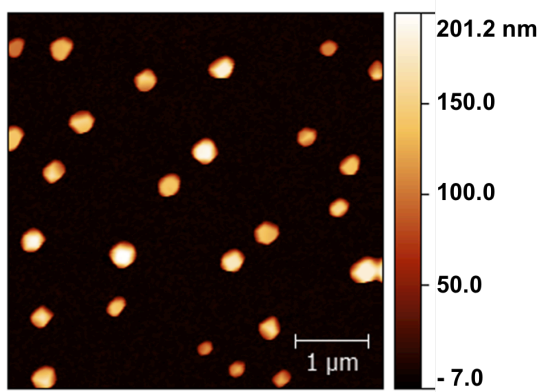

**Fig. (6).** Atomic force microscopy topographic height images of A172-derived exosomes precipitated by Exoquick. The exosomes appear as circular structures with an average diameter of about 70 nm. Courtesy of F. Caponnetto.

The most common technique for size-based separation of macromolecules is chromatography [101]. The biggest advantage of this approach is that the relatively small shearing forces do not affect biological particles thus maintaining the structure of vesicles intact. Another method based on size for exosomes isolation is ultrafiltration: this approach allows the separation of exosomes from proteins and other macromolecules with a higher and faster recovery of particles smaller than 100 nm, as compared to ultracentrifugation [101]. A micropillar porous silicon-ciliated structure has also been described for the isolation of 40-100 nm exosomes even if it hasn't been tested yet with clinical samples [102].

However, all these approaches are not able to isolate specific exosome subpopulations. This can be done by taking advantage of immunoaffinity strategies. For example, tumor-derived exosomes can be enriched from biological fluids by using beads functionalized with antibodies anti-EpCAM, anti HER2 or anti-PLAP, [103]. However, this method cannot be applied for large volumes of sample and in some cases it has caused a loss of exosome functional activity.

A sophisticated antibody-based method for exosomes enrichment is the Dynabeads technology. Superparamagnetic polystyrene beads are used to capture exosomal CD63 protein. This system requires some washing steps to remove non-specific antibody binding and this potentially affects the rate of particle recovery [104]. An ELISA-based method has also been developed for the detection of exosomes, in supports functionalized with anti-CD9 or anti-CD63 antibodies [105].

### 3.4. Microfluidic platforms

The increased interest in biological properties and roles of exosomes and their possible implications for the clinical practice has supported the growth of new microfluidic technologies for their study. The optimal platform should allow high throughput isolation of specific, pure populations of exosomes suitable for downstream analyses. The optimized platforms can be categorized in three main classes on the basis of the traditional strategy adopted for their capture: the purification, in fact, can be performed exploiting the trapping on particular porous surfaces, the sieving or the immunoaffinity separation.

*Porous surfaces*. A method based on the entrapment of exosomes within a porous surface has been described to be highly selective in the range of 40-100nm demonstrating a recovery rate of about 50% [102]. Downstream analysis of their cargo has not been performed and the system is not validated for clinical samples.

*Sieving*. An effective approach to collect exosomes with high purity is the sieving of biofluids. This non-selective isolation method takes the advantage of pressure driven filtration or electrophoresis on a membrane to collect exosomes. This approach has been used in whole blood and even if a low recovery rate has been demonstrated, it was capable of detecting sufficient material to be analyzed by RT-PCR and Western Blot [106].

*Immunochips*. As already mentioned, an immunological detection is currently the only way to perform a type-specific exosome enrichment and analysis. A new microfluidic system, described by Dudani et al., based on inertial focusing is able to enrich functionalized polystyrene beads at high speed, thus allowing exosomes characterization. The main drawback of this method is the long incubation time (four hours) for the binding of exosomes to the beads; this system has been tested in vitro with exosomes derived from cancer cell line and spiked into healthy blood. Downstream analysis of their cargo hasn't been performed yet [107].

Another example of a microfluidic device based on immunoaffinity is the Immuno-chip system that was developed to capture CD63 positive exosomes from biological fluids. Thanks to a planar structure with herringbone engravings, exosomes are trapped into the device and, after lysis, they are available for further investigations, including genomics and transcriptomics. Using the Immuno-chip for collecting exosomes from a non small-cell lung cancer (NSCLC) patient's blood sample the quality and the quantity of total RNA extracted from exosomes was suitable for downstream evaluations [108]. Using the same principle, another device, ExoChip, allows the enrichment and quantification of exosomes by an automatic fluorescence-based detection in a plate reader. Once again, the incubation time is crucial for the interaction between the exosomes and the functionalized CD63 surface. ExoChip has been used to isolate exosomes from healthy and pancreatic cancer patients' sera, making this method suitable for diagnosis and screening of human cancer patients through the analysis of their cargo of miRNAs. An important advantage of these methods based on immunoaffinity is that they are fast and with a high throughput [109,110].

Another microfluidic device, able of on-chip isolation, demonstrated interesting results on patients: the simplicity and speed of the method –requiring premixing of antibodies in plasma and exosomes lysis, protein immuno-precipitation and a fluorescence detection assay for quantification-demonstrated a recovery of exosomes 100 times lower than ultracentrifugation, but allowing the researchers to discriminate NSCLC patients from healthy volunteers [110].

Other devices are being developed to improve actual limits of immuno-based systems, namely the absence of multiplexing and rapid quantification: a notable example is nPLEX that allows a simultaneous detection of multiple markers. In a first generation system it was able to detect 36 different proteins thanks to 12 microfluidic channels coupled with 3 assays each, while a second-generation prototype should be able to detect a combination of up to 1089 (33x33) proteins. The technology behind this device, the surface plasmon resonance, is highly sensitive, thus allowing precise quantification in real-time. A study on 10 healthy volunteers and 20 ovarian cancer patients demonstrated its potential use as a diagnostic tool: in the study ascitic fluid was used but virtually all biological fluids can be tested if an adequate pre-analytic phase is implemented [111].

## 4. 3D CULTURE MODELS FOR CELL-CELL INTERACTION ANALYSES

### 4.1. Tumor microenvironment

Solid tumors are complex organ-like entities that consist not only of malignant cells but also of many non-transformed cell types, soluble factors, signaling molecules and extracellular matrix (ECM) components that altogether compose the tumor microenvironment (TME) [112]. Malignant cells establish myriad interactions with neighboring non-malignant cells and ECM by different pathways that include complex and dynamic network of cytokines, chemokines, growth factors, and inflammatory and matrix remodeling enzymes [113]. Major non-malignant cell types that are found in the TME include endothelial cells (ECs) of the blood and lymphatic circulation, cancer-associated fibroblasts (CAFs), mesenchymal stem cells (MSC), and a variety of different immune cells including lymphocytes and tumor-associated macrophages (TAMs) [114]. It has became widely accepted that the non-malignant cellular components residing in the stromal microenvironment of tumors not only influence the tumor proliferation and metastasis but are also capable of altering the response of tumors to diverse therapeutics [63,115,116]. Growing evidence indicates that different subsets of T cells are critical for tumor generation and maintenance [117,118] and are the subject of recent advances in the development of novel immunotherapies [119]. The alternative activation of macrophages and their ability to promote tumor development and progression have also been well-studied [120]. Other studies have shown how myofibroblasts and MSCs derived from the bone marrow directly support cancer stem cells (CSCs) by creating a favorable niche and facilitating tumor progression [121]. Another group of cells that have a major role in tumorigenesis includes endothelial cells and pericytes, which play a key role in vascular functionality and angiogenesis, as well as in regulating cancer cell dissemination [122]. The ECM, which provides physical scaffolding for the cellular constituents and initiates crucial biochemical and biomechanical signals that are required for tissue morphogenesis, differentiation and homeostasis, is commonly deregulated and becomes disorganized in tumors. It has been shown that abnormal ECM affects cancer progression by directly promoting cancer cell transformation and metastasis [123,124]. Importantly, ECM anomalies also deregulate behavior of stromal cells and thus lead to generation of a tumorigenic microenvironment that further facilitates cancer progression. For example, many ECM fragments, including endostatin, tumstatin, canstatin, arresten, and hexastatin [125], are likely to collaborate with other pro- or antiangiogenic factors, including vascular endothelial growth factor (VEGF), to determine where to initiate vascular branching and the final branch pattern. Recent studies showed that increased ECM stiffness can promote integrin-mediated adhesion complex assembly and activate T cells [126–129]. Therefore, a deep analysis of the complex interactions of tumor cells with their surrounding microenvironment is fundamental in developing effective anti-cancer therapeutics for a superior or complete response on patient treatments.

### 4.2. Limitations of *in vitro* 2D monolayer models

To date, adherent two-dimensional (2D) cell monolayers are still largely used for a wide range of cell-based assays in both basic and clinical cancer research, however experimental data obtained in 2D are rarely translated *in vivo* [130–132]. This is primarily a consequence of the inability of monolayer cultures to recapitulate fundamental characteristics of tumors including cell–cell and cell–matrix interactions as well as hypoxia, dormancy and anti-apoptotic features. Drugs delivered to cells cultured on flat and rigid substrates reach cells without encountering barriers or without exposure to critical oxygen and nutrient gradients that can reduce their diffusion profile or inhibit their action. Accordingly, compared to *in vivo* conditions, cells in 2D generally exhibit a higher sensitivity to some therapeutic agents [133,134]. Moreover, 2D cultures conditions, including subculturing and long-term passages, may alter gene expression and phenotype [135]. Some alterations may sustain for a long time resulting in the selection of population of cancer cells that significantly vary from the cancer cells found in the patient's tumor [136–138]. *In vitro* 2D cultures also influence the shape of the cells that usually grow with a more flattened and stretched morphology in comparison to their counterparts *in vivo* conditions. Such abnormal cell morphology is known to influence numerous cellular processes including cell proliferation, differentiation, and apoptosis that in turn affect cell response to anticancer therapy [139–141]. Cancer cells grown on stiff 2D substrates before inoculation in animal models may also fail to create xenograft animal models that recapitulate the exact phenotypes of human tumors. For instance, Leung M *et al.* [142], reported that cancer cells cultivated in three-dimensional chitosan-alginate scaffolds before implantation in mice accelerated tumor growth and promoted angiogenesis, compared to 2D cultured cells. Other findings showed that injection in mice of cells previously cultured in 3D fibrin gels led to the formation of solid tumors more efficiently than cancer cells selected using conventional 2D cultures [143]. Three-dimensional (3D) cultures with appropriately engineered cell culture environment can thus overcome a number of limitations of 2D cultures aiding our understanding of cancer biology and improving the predictive accuracy of the drug discovery process prior to testing in animal models and ultimately clinical trials.

**4.3. 3D culture models**

The ideal 3D tumor model must closely recapitulate the conditions found *in vivo* by supporting the growth of multiple cell types while providing the appropriate matrix environment and mass transport. It should allow monitoring and adjusting the hypoxia levels as well as reporting on the release of angiogenic factors from the cancer cells in response to a specific stimulus. The growing numbers of publications in the field of "3D cell culture" systems reflects the rapid advancement in technology and biomaterials which are providing new opportunities for creating pathologically-relevant 3D models that imitate tumor morphology [144], protein expression [145,146], tumor-stroma interactions [147], as well as chemical and biological gradients [148]. Methods for assembly of 3D cell cultures are generally classified as scaffold-free and scaffold-based culture systems, each of these systems possessing advantages and disadvantages. Therefore, the most appropriate platform must be selected with respect to its suitability to the individual needs of the user [149]. In the following, we describe the main properties and applications of 3D tumor models based on spheroids, hydrogels, and fibrous-based scaffolds (Fig. **7**). Notably, numerous microfluidic devices have been designed for integrating gel-based systems and spheroid-based systems into 3D microfluidic tumor models. The use of microfluidics allows for the creation of 3D platforms with more complex and well-controlled in-vivo like 3D dynamic environment (e.g., control of oxygen, nutrients and cellular gradients) as well as for high-throughput drug screening. For a detailed description of the advances of microfluidic technology in 3D cell cultures the readers can refer to the following reviews [150,151].

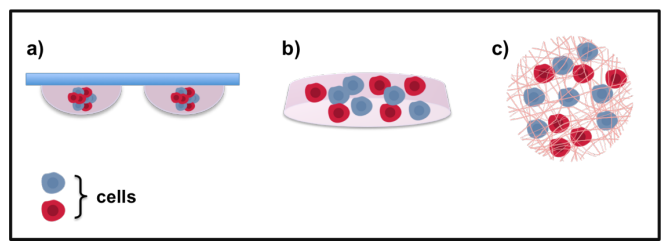

**Fig. 7.** Schematic illustration of **a)** spheroids, **b)** hydrogel, and **c)** fibrous-based scaffold. Each system can support growth and rearrangement of more than one cell type and allows for exchange of oxygen, nutrients and bioactive molecules.

*4.3.1 Spheroids*

Spheroids are aggregates of cells made of a core of quiescent or hypoxic cells surrounded by an outer layer of viable and proliferating cells [152]. It is well accepted that 3D spheroids are reliable models of *in vivo* solid tumors since they not only recapitulate cell–cell and cell–matrix interactions between tumor cells and the microenvironment [153,154], but also present natural transport properties [152] that result in oxygen and nutrients gradients reminiscent of tumors. The shape and size of spheroids play a crucial role since they are correlated with cell function, drug penetration and transport. For instance, normal prostate epithelial cells differentiate into well-polarized hollow spheroids, whereas malignant cells typically form atypical spheroids with disorganized architecture, as demonstrated most prominently for breast cancers [135,155], Concerning the size, spheroids with size ranging from 200 μm to 500 μm are usually sufficiently packed to recapitulate cell–cell and cell–matrix interactions and are large enough to develop gradients of oxygen, nutrients, and catabolites [153]. A number of techniques have been developed to create 3D spheroids, such as hanging drop technique [156,157], continuous agitation of suspension culture in a rotary cell culture vessel[158] or a spinner flask [159], preparation of cell repulsive substrates [160], and entrapment within biologically inert 3D hydrogel matrices [161,162]. 3D spheroids have been proposed for studies of cell function, drug testing, tumor angiogenesis, as well as for the study of tumor–immune cell interactions [153,154,163,164]. Spheroids have also been used to cultivate cancer stem cells [165–167]. As an example, spheroid cultures of human gingiva-derived mesenchymal stem cells showed better therapeutic efficacy than their adherent cells in reversing body weight loss and promoting the regeneration of disrupted epithelial lining of the mucositic tongues [166]. Although the numerous advances in the production and utilization of spheroids, there is still a lack of standardized procedures for producing spheroids of uniform size in a reliable, sustainable and reproducible manner. Other challenges involve forming stable culture of spheroids from a small number of cells and designing standardized assays for rapid analysis of cellular responses *in situ* making them compatible with readouts associated with drug delivery and efficacy testing.

*4.3.2 Hydrogels*

Hydrogels are water-swollen 3D networks of cross-linked polymer chains with highly tunable biophysical and biochemical properties. They can be prepared from either natural polymers such as fibrin [168,169], hyaluronic acid [170,171], collagen [172], alginate [173], and chitosan [174] or synthetic polymers such as poly(ethylene glycol) [175] (PEG) and PEG derivatives, poly(lactic acid) [176] and poly(vinyl alcohol). Hybrid composite hydrogels consisting of synthetic and natural polymers have also been prepared to improve the mechanical and biological properties of the scaffolds. Usually, hydrogels are made by quickly mixing the liquid precursor solutions with a cell suspension, followed by rapid addition of a gelling or cross-linking agent. The gelling process must be fast to avoid cells from sinking to the bottom of the scaffold. The methods for preparing hydrogels have been reviewed elsewhere [177–179]. Hydrogels represent a valid alternative platform for 3D cell culture because they closely resemble the physical characteristics of native extracellular matrix [180] and show mechanical properties similar to natural living tissues. Moreover, they provide a 3D culture micro-environment that can be easily tuned by simply varying the nature of the polymers and by incorporating bioactive molecules (e.g., peptides) to allow growth of specific cell types [181,182]. For instance, PEG hydrogels modified with an integrin-binding RGD peptide and matrix metalloproteinase-degradable peptide crosslinkers were shown to promote 3D epithelial morphogenesis of lung adenocarcinoma cells similarly to 3D culture in Matrigel [183]. Another appealing feature of hydrogels as 3D scaffolds is their ability to encapsulate and release bioactive agents, including regulatory factors [184]. Because of their versatile and unique properties, 3D hydrogels have proven useful for various studies including drug screening [185,186], study of angiogenesis [187,188], and study of mechanical influence on cell behavior [189]. Despite their great versatility, hydrogels have also some inherent drawbacks including the use of ultraviolet irradiation which may have adverse effects on cellular metabolic activity [190], and the possible limited diffusion of nutrients across gels that limit culture of cells for long times [182]. Another limitation is that it is difficult to introduce multiple cell lines at various time points once the hydrogel is formed making challenging to study intercellular signaling.

*4.3.3 Fibrous-based scaffolds*

Fibrous scaffolds consist of fibrous matrices that provide a 3D space in which cells can grow recreating natural tissue-like structures. A number of methods are available for producing 3D fibrous scaffolds. Electrospinning has emerged as simple and scalable one-step method for the fabrication of dense fiber meshworks that closely resemble the fibrillar components of the native ECM [191]. The fabrication procedure involves the use of an electric field to propel a thin jet of polymer from an electrically biased syringe to a target grounded surface [192,193]. Depending on the parameters of the electrospinning setup and on the nature of the polymers, fibers with diameter ranging from tens of nanometers to several microns can be obtained. By using two or more jetting polymers it is also possible to produce heterogeneous fiber mats. Numerous types of natural and synthetic polymers have been successfully employed, either alone or in combination, to produce micro- or nanofibrous scaffolds. FDA-approved polymers include poly(glycolic acid), poly(L-lactic acid), poly(D,L-lactic acid), poly(D,L-lactic-co-glycolic acid 50:50), poly(D,L-lactic-co-glycolic acid 85:15), and poly(ε-caprolactone) [194]. Fibrous-based scaffolds produced by electrospinning possess unique physical characteristics such as high surface area-volume ratio and improved mechanical strength [191,195]. Another advantageous property is their ability to support cell-ECM interactions [196,197] and cell-matrix adhesion making them ideal matrices to study cell-signaling pathways. Electrospun fibers are also used as ideal platform to create artificial 3D stem cell niche with defined topographical and biochemical cues that can impact the local regulation of stem cells [198–200]. By varying the electrospinning procedure, aligned and randomly oriented fibers can be obtained [192,201]. Aligned fibrous scaffolds offer interesting opportunities to study alignment, migration and directionality of neural cells [202,203]. Moreover, thanks to the high versatility and simplicity of the electrospinning procedure, multifunctional composite electrospun scaffolds can be obtained by functionalizing the fibers with a large variety of nanomaterials [204–206] (e.g., nanoparticles, sensors) and bioactive molecules [207,208] (e.g., drugs, growth factors, proteins, peptides and DNA) for programmable multi-agent delivery and sensing applications. Taken together, these features make elctrospun fibrous scaffolds particularly advantageous for 3D cell culture studies. However, while nanofiber matrixes provide excellent conditions for cell attachment and spreading on the surface of the scaffold, they prevent efficient cell migration into the fiber mats and thus limit study of cell invasion inside the engineered constructs.

**CONCLUSION**

The urgent need to develop more effective treatments for cancer patients and to predict therapeutic response is the rationale for the development of novel research tools for studying cancer cells at single cells level and within the tumor microenvironment.

We have seen that, despite a tremendous rise in the number of microfluidic technologies for the detection of CTCs, none of them managed to obtain FDA approval. New microfluidic techniques must prove their clinical validity beyond cancer cell lines, in the tough arena of clinical trials, which require a level of robustness and standardization of the assay that often goes far beyond academic published results.

Affinity-based methods are the most robust ones due to the long history of monoclonal antibody production and usage in research environments, but with downsides like "a priori" target selection, the unknown dynamic biology of cancer protein expression and cross-reactions. "Label-free" techniques are promising but yet unvalidated methods, which might bring high throughput (size, charge-based methods) high sensitivity/specificity (metabolism-based method) and low-cost to the panorama. Most likely, an integration of these different methods in different steps of enrichment and detection can bring a solution providing required sensitivity, specificity, cost and throughput to the clinic.

Regarding exosomes, microfluidic technologies would greatly help in isolating this extracellular vesicle subset whose size makes extremely challenging their quantification and characterization by fast, sensitive, specific and high throughput assays. However, this field is still facing many biological and technical issues. Biologically, extracellular vesicles are constituted by extremely heterogeneous and overlapping subsets. Therefore, each exosome isolation method, either physical or affinity-based, is mainly enriching in rather than purifying exosomes. Moreover, exosome composition is changing depending on many factors, including cell types and clinical conditions. Thus, affinity-based methods can be affected by a "a priori" selection. From a technical point of view, it has to be taken into consideration that many exosomes enrichment procedures are indeed co-precipitating or co-isolating non-exosomal components, such as circulating proteins or nucleic acids, which can interfere with down-stream analyses. Finally, it is becoming apparent that different isolation methods can enrich in functionally different exosome subsets [209], making not only desirable but also necessary future works that involve a direct comparison of different methods.

Significant advances in nanotechnology and biomaterials over the past decades allowed the design of 3D platforms suitable for studying molecular and genetic mechanisms that drive cancer growth and progression under *in vivo*-like conditions. It has been widely recognized that, compared to 2D cell-cultures, 3D cell cultures provide a better environment for studying cell function (e.g., aggregation, migration, proliferation and invasion), efficacy of drugs and drug action under *in vivo*-like conditions. However, further research is needed to develop 3D models that can fully replace *in vivo* models by reproducing the vast heterogeneity and complexity of tumor microenvironment, tumor growth and metastases cascades. It is foreseen that the collaboration between clinicians, cancer biologists, chemists, physicians, bioengineers and mathematicians will accelerate the coupling of different technologies for creation of 3D tumor models in which multiple aspects of tumors can be analyzed in parallel and quantified.

## CONFLICT OF INTEREST

The authors confirm that this article content has no conflict of interest.

## ACKNOWLEDGEMENTS

This work has been partially founded by the ERC projects "QUIDPROQUO" (Advanced Grant No. 269051) and "A CACTUS" (PoC Grant No 640955). The national projects "Molecular Nanotechnologies for Health and Environment" (MAAT, PON02_00563_3316357), "Find New Molecular and Cellular targets against cancer" (FIERCE, FIRB No. RBAP11Z4Z9), "Nanotechnological approaches for tumor theragnostic" (FIRB No. RBAP11ETKA_007), "Application of Advanced Nanotechnology in the Development of Cancer Diagnostics Tools" (AIRC 5 per mille Special program 2011 No. 12214) and "Beyond-Nano" (PONa3_00363) are also acknowledged for financial support.